\documentclass[preprint, 12pt]{elsarticle}
\usepackage{amssymb}
\usepackage{amsmath}
\usepackage{url}
\usepackage[margin=2.5cm]{geometry}

\usepackage{siunitx}

\begin{document}

\newpageafter{abstract}

\title{Anisotropic swelling due to hydration constrains anisotropic elasticity in biomaterial fibers} 
\author[1]{Xander A. Gouws}
\author[1]{Ana Mastnak}
\author[1]{Laurent Kreplak}
\author[1]{Andrew D. Rutenberg}
\ead{adr@dal.ca}

\affiliation[1]{organization={Department of Physics and Atmospheric Science},
            addressline={Dalhousie University}, 
            city={Halifax},
            postcode={B3H 4R2}, 
            state={Nova Scotia},
            country={Canada}}

\begin{abstract}
Naturally occurring protein fibers often undergo anisotropic swelling when hydrated. Within a tendon, a hydrated collagen fibril's radius expands by 40\% but its length only increases by 5\%. The same effect, with a similar relative magnitude, is observed for single hair shafts.  Fiber hydration is known to affect elastic properties. Here we show that \emph{anisotropic} swelling constrains the anisotropic linear elastic properties of fibers. First we show, using data from disparate previously reported studies, that anisotropic swelling can be described as an approximately linear function of water content. Then, under the observation that the elastic energy of swelling can be minimized by the anisotropic shape, we relate swelling anisotropy to elastic anisotropy -- assuming radial (transverse) symmetry within a cylindrical geometry. We find an upper bound for the commonly measured axial Poisson ratio $\nu_{zx}<1/2$. This is significantly below  recently estimated values for collagen fibrils extracted from tissue-level measurements, but is consistent with both single hair shaft and single collagen fibril mechanical and hydration studies. Using $\nu_{zx}$, we can then constrain the product $\gamma \equiv (1-\nu_{xy}) E_z/E_x$ --- where $\nu_{xy}$ is the seldom measured transverse Poisson ratio and $E_z/E_x$ is the ratio of axial to radial Young's moduli. 
\end{abstract}
\date{\today} 
\maketitle

\section{Introduction} 
Many biomaterials absorb water in response to changes in the ambient humidity. Leonardo da Vinci used this effect to make the first hygrometer, or humidity sensor, in 1481 using the changing mass of cotton with humidity. In 1783, hair hygrometers were developed that used the changing length of hair shafts with humidity~\cite{Korotcenkov:2019}, and were widely used until the last century. More generally, any biomaterial or biotextile will have changing water content with humidity (see, e.g.,~\cite{Yin:2023}). The increasing development and use of biomaterials and biotextiles makes the systematic effects of hydration important to characterize.

For relatively well-studied biological fibers, such as hair shafts and collagen fibrils, the effects of hydration are dramatic. Mechanical properties of hair shafts depend strongly on water content~\cite{Popescu:2007}, as do modes of mechanical failure~\cite{Kamath:1982}. The Young's modulus of collagen fibrils varies by orders of magnitude between dry and aqueous conditions~\cite{Andriotis:2018}; similar scale changes are seen in bending~\cite{Yang:2008} or indentation~\cite{Andriotis:2018, Grant:2008} and in keratin appendages \cite{Johnson:2017}. Furthermore, mechanical effects of hydrated fibers are strongly affected by solution conditions --- reflecting different water contents~\cite{Grant:2009, Masic:2015, Haverkamp:2022}.

Both hair shafts and collagen fibrils are strongly anisotropic materials, with fibril-forming alpha-keratins or fibril-forming collagen molecules predominantly aligned with the cylindrical axis of individual hair shafts or collagen fibrils, respectively. Substantial anisotropy is also observed with hydration effects. When hydrated, a collagen fibril's radius may expand by~40\% while its length only increases by~5\%~\cite{Haverkamp:2022}. Similarly, when a hair shaft is hydrated, its radius expands by~14\%, but its length expands by only~2\%~\cite{White:1947, Stam:1952}. 

Substantial elastic anisotropy is also observed in these systems. Studies of single collagen fibrils exhibit ratios of axial to radial stiffness greater than $20$ \cite{Andriotis:2018, Andriotis:2023}. A study of human hair shafts shows stiffness ratios as large as $2$ \cite{Breakspear:2018}. How this mechanical anisotropy may relate to the swelling anisotropy has not been examined, though both swelling and mechanical anisotropies should ultimately arise from the anisotropic structures of these biomaterial fibers.

Our goal in this paper is to simplify, summarize, and compare the anisotropic phenomenology due to hydration in both hair and collagen fibers. Since the water content of protein fibers can vary by osmolarity even at a fixed relative humidity~\cite{Grant:2009, Masic:2015, Haverkamp:2022}, we investigate how shape depends on water content. We obtain a simple phenomenological model for this relationship. We use this model to constrain anisotropic linear elastic models --- independently of any microscopic structural model. Since these elastic models are coarse-grained, we also obtain an effective coarse-grained mechanism for anisotropic swelling in biomaterial fibers. We expect this elastic picture of swelling to generically apply for small strains, where linear elasticity applies. 


\section{Methods and Results} 
\begin{figure}[t]
    \centering
    \includegraphics[scale=0.5]{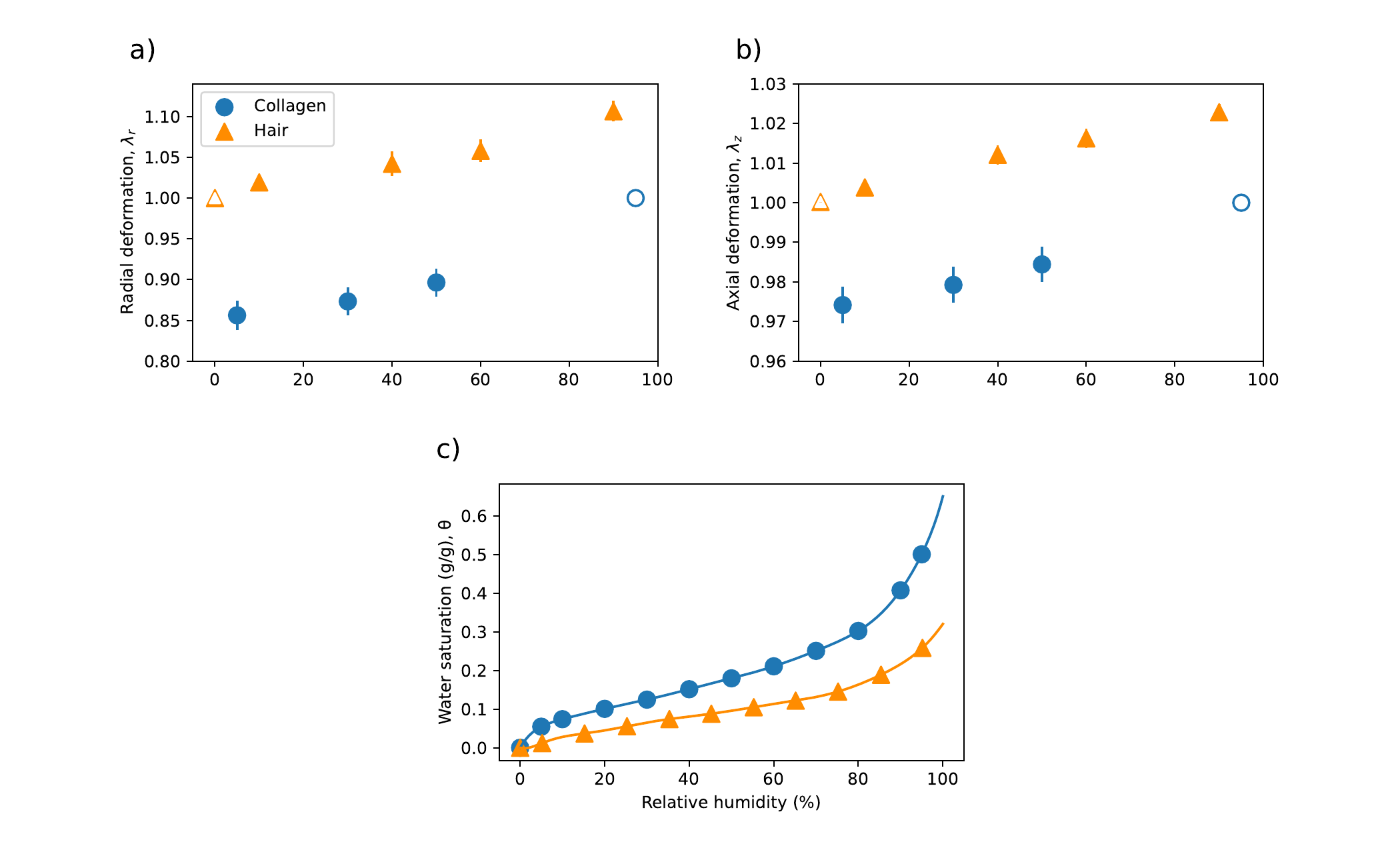}
    \caption{Previously reported hydration data. a) radial deformation of hair shafts (``hair'') \protect\cite{Stam:1952} and collagen fibrils (``collagen'') \cite{Masic:2015}, b) axial deformation of hair \cite{Stam:1952} and collagen \cite{Masic:2015}, c) water saturation (g water/g material) for hair \cite{Barba:2010} and collagen \cite{Bull:1944}, with interpolating lines fit using the Guggenheim–Anderson–de Boer (GAB) model from \cite{Barba:2010}. For hair, the hydration curve is shown. All are vs. relative humidity ($\%$). Standard errors are indicated. Unfilled points indicate the baseline states.}
    \label{fig:raw-data}
\end{figure}

Relative dimensional changes induced by variations in relative humidity have been previously experimentally characterized.  We have extracted digitized collagen fibril data~\cite{Masic:2015}, and used raw hair shaft hydration measurements~\cite{Stam:1952}. Fig.~\ref{fig:raw-data}a shows the radial deformation
\begin{equation}\label{eq:define-lambda-r}
    \lambda_r = R/R_0,
\end{equation}
for both collagen fibrils (``collagen'', blue circles in all figures) and hair shafts (``hair'', orange triangles) vs. relative humidity, where $R$ is the cylindrical radius and $R_0$ is the baseline radius at the relative humidity indicated by the unfilled points. All error bars in this paper are standard errors of the means (``standard errors''), corresponding to one standard deviation. Fig.~\ref{fig:raw-data}b shows the corresponding axial deformations
\begin{equation}\label{eq:define-lambda-z}
    \lambda_z = L/L_0,
\end{equation}
where $L$ is the axial length and $L_0$ is the baseline length. 

\begin{figure}[h]
    \centering
    \includegraphics[scale=0.45]{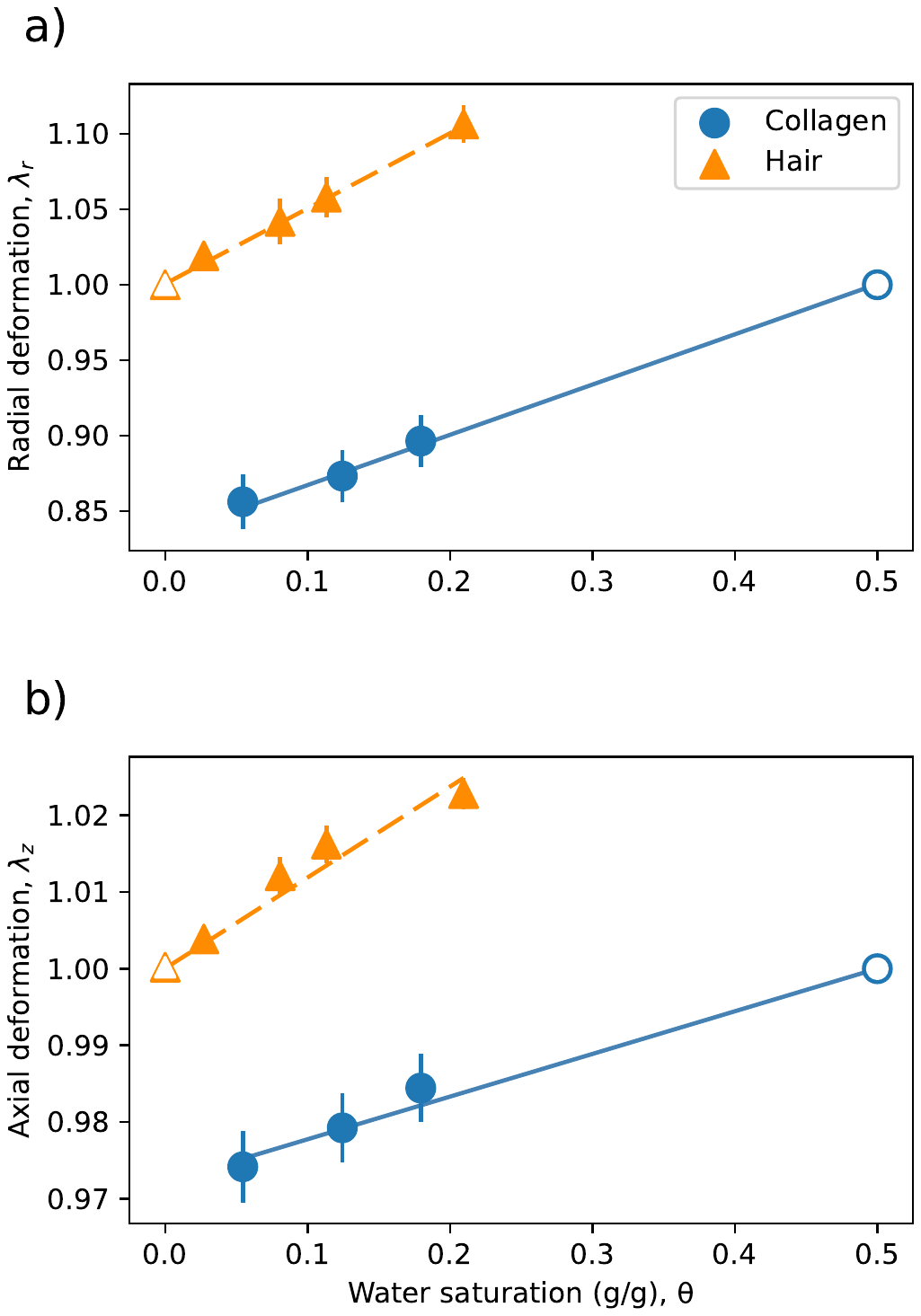}
    \caption{Deformations vs water saturation using the data from Fig.~\ref{fig:raw-data}. a) Radial deformation, $\lambda_r$ vs. saturation $\theta$ (grams of water absorbed per gram of dry mass). b) Axial deformation $\lambda_z$ vs. $\theta$. Unfilled points indicate the baseline states.  Coloured lines are linearized from fits in Fig.~\ref{fig:saturation vs alpha and beta}.}
    \label{fig:saturation vs deformations}
\end{figure}

\begin{figure}[h]
    \centering
    \includegraphics[scale=0.45]{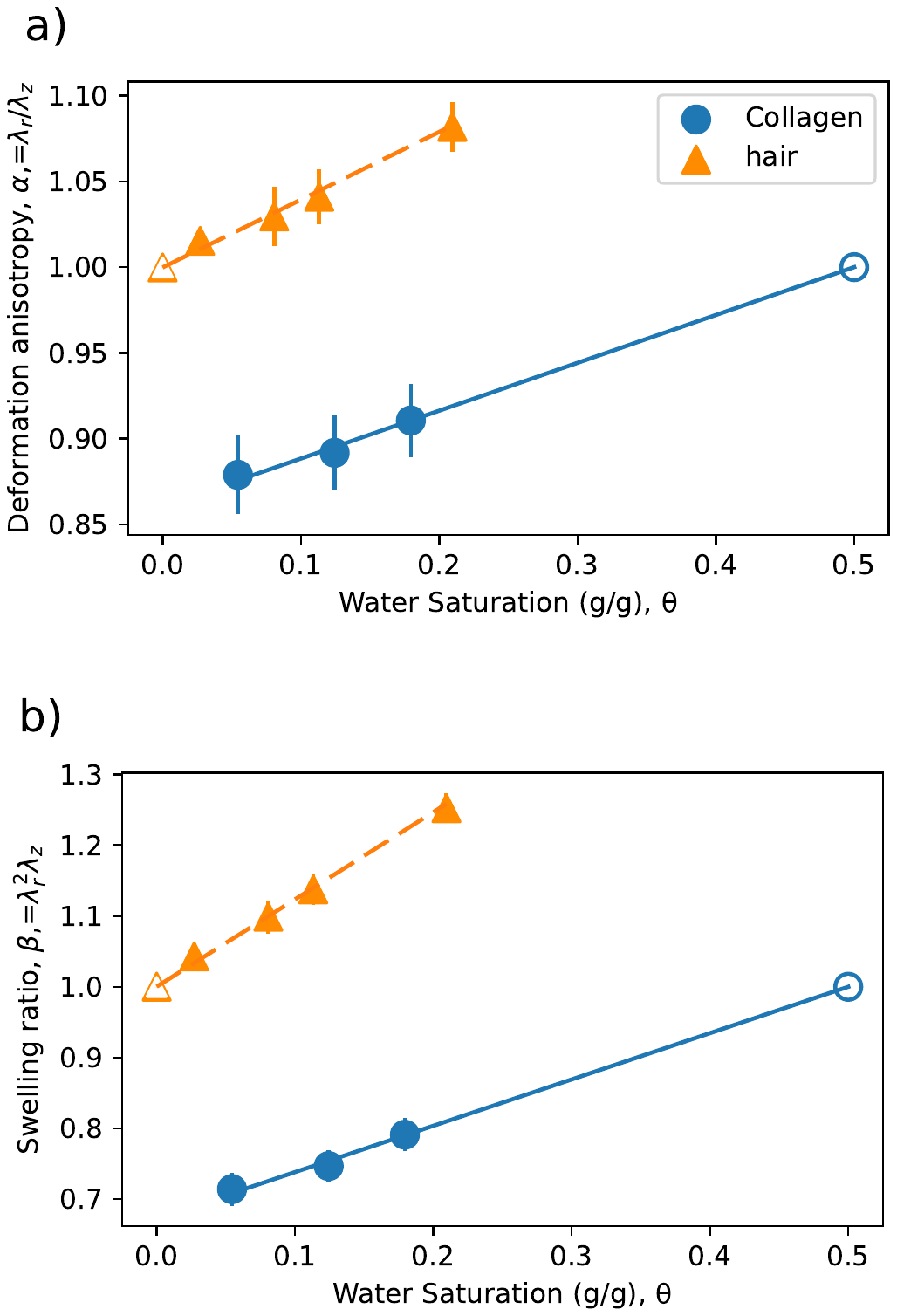}
    \caption{Geometric deformations vs water saturation. a) Deformation anisotropy $\alpha = \lambda_r/\lambda_z$ vs. saturation $\theta$. b) Swelling ratio $\beta = \lambda_r^2 \lambda_z$ vs. $\theta$.  Lines are linear best fits.}
    \label{fig:saturation vs alpha and beta}
\end{figure}

We are interested in how the water content of hair and collagen affects their shapes and structures. Independent studies have examined the relationship between water content and humidity both in hair~\cite{Barba:2010} and in collagen~\cite{Bull:1944}, as shown in Fig.~\ref{fig:raw-data}c. These experiments report saturation $\theta$ as the grams of water absorbed per gram of protein dry mass. Errors were not reported for these saturation studies. How deformations change with saturation $\theta$ has not been previously reported, but can be extracted by using the deformation studies exploring ambient humidity~\cite{Stam:1952, Masic:2015}. 

The experimental data used here \cite{Stam:1952, Masic:2015, Barba:2010, Bull:1944} are from different groups, with different samples, and with different  techniques. For the hair swelling studies \cite{Stam:1952}, human hair strands with dry diameters between $40-60 \mu m$ and with cross-sectional elliptical eccentricity less than 1.1 were used. For the hair water content studies \cite{Barba:2010}, untreated human hair was used. For the collagen water content studies \cite{Bull:1944}, non-fibrillar collagen extracted from `hide' was used. For the collagen swelling studies \cite{Masic:2015}, radial and axial strain of fibrils during swelling was extracted from x-ray studies of unstressed rat tail tendon. These experiments have not been quantitatively replicated by other groups and do not include  estimates of systematic errors, so that the errors shown may represent underestimates of the actual experimental errors. 

In Fig.~\ref{fig:saturation vs  deformations}, we re-plot the deformation data from Fig.~\ref{fig:raw-data} as a function of saturation. We have interpolated the saturation data of hair as needed, using the theoretical function given by \cite{Barba:2010} and shown in Fig.~\ref{fig:raw-data}c. Note that the hair study used a dehydrated baseline (at $\theta_0 \approx 0$) and explored the effects of increased hydration while the collagen study used a hydrated baseline ($\theta_0 \approx 0.5$) and then dehydrated the samples. We see that the  deformations vary approximately linearly with saturation. Since $\lambda \approx 1$, deformations to any low power  are also approximately linear with respect to saturation.

To investigate how shape changes with water content, we define the deformation anisotropy
\begin{equation}
    \alpha \equiv {\lambda_r}/{\lambda_z},
\end{equation}
which is a measure of how anisotropic the deformations are. We also define a geometric measure of water content, the  swelling ratio
\begin{equation}
    \beta \equiv \lambda_r^2 \lambda_z = V/V_0,
\end{equation}
which is how much the volume ($V$) has changed with respect to the baseline volume ($V_0$).  Our phenomenological model is that both $\alpha$ and $\beta$ are linear functions of the saturation $\theta$ (i.e. the water content):
\begin{equation}\label{eq:alpha vs theta}
    \alpha = 1+A (\theta - \theta_0)
\end{equation}
and
\begin{equation}\label{eq:beta vs theta}
    \beta = 1+B(\theta - \theta_0). 
\end{equation}
The initial saturation $\theta_0$ is the saturation at which the baseline lengths $R_0$ and $L_0$ have been measured, where $\lambda=\alpha=\beta=1$. We generically expect $B>0$ if added water leads to added volume, but we have no such constraint on $A$.

In Fig.~\ref{fig:saturation vs alpha and beta}, $\alpha$ and $\beta$ have been plotted as functions of the saturation together with best-fit lines from our linear model Eqns.~\ref{eq:alpha vs theta} and \ref{eq:beta vs theta}. For hair and collagen, the best-fit slope parameters are
\begin{align}
    A_\text{collagen} &= 0.28\phantom{0} \pm 0.03, \label{eq:slope_start}\\
    B_\text{collagen} &= 0.65\phantom{0} \pm 0.06,\\
    A_\text{hair} &= 0.40\phantom{0} \pm 0.05, \text{ and}\\
    B_\text{hair} &= 1.23\phantom{0} \pm 0.08, 
    \label{eq:slope_end}
\end{align}
where errors are standard errors (representing one standard deviation).
These parameters have been determined using Deming regression with the \textit{SciPy} version 1.13.1 orthogonal distance regression subpackage. With no errors in $\theta$, this performs ordinary least squares fitting whilst accounting for uncertainty in the data.  

\begin{figure}[t]
    \centering
    \includegraphics[scale=0.7]{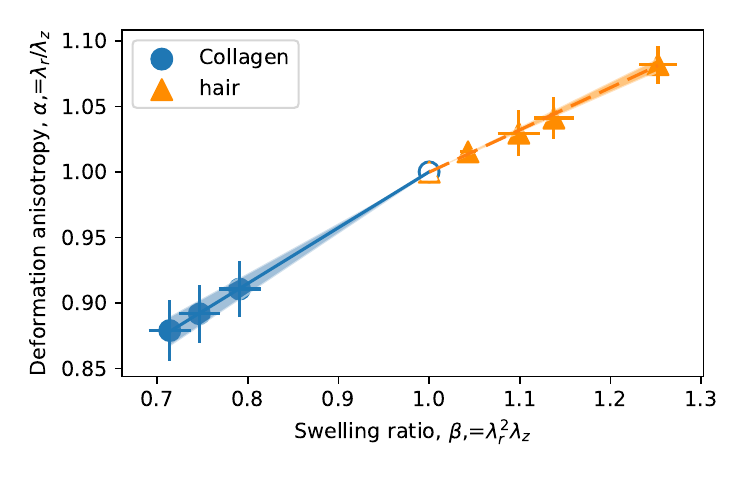}
    \caption{Deformation anisotropy $\alpha = \lambda_r/\lambda_z$ vs. swelling ratio $\beta = \lambda_r^2\lambda_z = V/V_0$, for both hair and collagen as indicated. Unfilled points indicate the baseline state at $\alpha_0=\beta_0=1$. The lines are linear best-fits, while the shaded regions indicate standard errors of the fit.}
    \label{fig:alpha vs beta}
\end{figure}

We can plot $\alpha(\theta)$ and $\beta(\theta)$  against each other as a parametric function of the saturation $\theta$. This is shown in Fig.~\ref{fig:alpha vs beta}. Since the baseline point at $\alpha_0=\beta_0=1$ must be shared, we expect from Eqs.~\ref{eq:alpha vs theta} and \ref{eq:beta vs theta} that 
\begin{equation}
    \alpha - 1 = m (\beta - 1), 
\end{equation}
where $m$ is a slope parameter. While we have $m = A/B$, we refit $m$ to estimate standard errors using Deming regression. We obtain:  
\begin{align} \label{eq:m collagen}
    m_\text{collagen} &= 0.43 \pm 0.04, \text{ and}\\
    m_\text{hair} &= 0.32 \pm 0.03.
    \label{eq:m hair}
\end{align}
Values of $A$ and $B$ let us (at leading order in $\theta$) estimate $\lambda_r \simeq 1+(\theta-\theta_0)(A+B)/3$ and $\lambda_z \simeq 1+(\theta-\theta_0)(B-2A)/3$. These estimates are shown as straight lines in Fig.~\ref{fig:saturation vs deformations}. Note that since we expect that $\lambda_z$ grows with water content (along with $\lambda_r$), we should have $B>2A$ and so $m \simeq A/B <1/2$ --- as observed.

\section{Anisotropic elasticity}
Linear elastic theory allows us to explore the relationship between anisotropic swelling and anisotropic elastic properties for small volume changes. For a given water content (swelling ratio $\beta$), we will determine the anisotropic shape ratio $\alpha$ that minimizes the elastic strain energy. We will use this to relate $m$ to anisotropic elasticity.

Taking fibers as transversely isotropic orthotropic materials, the compliance matrix relating strain and stress is \cite{elastic-book-2}
\begin{equation}\label{eq:appendix-orthotropic-compliance}
    \begin{bmatrix}
        \varepsilon_{xx} \\
        \varepsilon_{yy} \\
        \varepsilon_{zz} \\
    \end{bmatrix} = 
    \begin{bmatrix}
        1/E_x & -\nu_{yx}/E_y & -\nu_{zx}/E_z  \\
        -\nu_{xy}/E_x & 1/E_y & -\nu_{zy}/E_z \\
        -\nu_{xz}/E_x & -\nu_{yz}/E_y & 1/E_z \\
    \end{bmatrix}
    \begin{bmatrix}
        \sigma_{xx} \\
        \sigma_{yy} \\
        \sigma_{zz} \\
    \end{bmatrix},
\end{equation}
where we have assumed zero shear. This relates the strains $\varepsilon_{ii}$ to the stresses $\sigma_{ii}$ through the Young's moduli $E_i$ and Poisson ratios $\nu_{ij}$, where the axial (longitudinal) direction is $z$ and the symmetric radial directions are $x$ and $y$. Using this radial symmetry, we have $E_x=E_y$, $\varepsilon_{xx}=\varepsilon_{yy}$, $\sigma_{xx}=\sigma_{yy}$, $\nu_{xy}=\nu_{yx}$, and $\nu_{xz}=\nu_{yz}$. Since the compliance matrix is also symmetric, we further have $\nu_{zx}/E_z=\nu_{xz}/E_x$. As a result, there are only four independent elastic parameters that are relevant: $E_x$, $E_z$, $\nu_{xy}$, and $\nu_{zx}$. 

Taking advantage of symmetries, and inverting the resulting $2 \times 2$ matrix, we obtain 
\begin{equation}\label{eq:stress-stiffness-tensor-strain}
    \begin{bmatrix}
        \sigma_{xx} \\
        \sigma_{zz}
    \end{bmatrix} = \frac{1}{1 - \nu_{xy} - 2 \nu_{xz} \nu_{zx}} \begin{bmatrix}
        E_x & \nu_{zx} E_x \\
        2\nu_{zx} E_x & (1 - \nu_{xy}) E_z
    \end{bmatrix}
    \begin{bmatrix}
        \varepsilon_{xx} \\
        \varepsilon_{zz}
    \end{bmatrix}.
\end{equation}
For small strains, the elastic strain energy density is
\begin{equation}
    u = \frac{1}{2}\left(\varepsilon_{xx}\sigma_{xx} + \varepsilon_{yy}\sigma_{yy} + \varepsilon_{zz}\sigma_{zz} \right) 
    = \varepsilon_{xx} \sigma_{xx} + \frac{1}{2} \varepsilon_{zz} \sigma_{zz}.
\end{equation}
Using Eq.~(\ref{eq:stress-stiffness-tensor-strain}) this gives
\begin{equation}\label{eq:strain-potential-energy-density}
    u = \frac{1}{1 - \nu_{xy} - 2 \nu_{xz}\nu_{zx}} \left[E_x\varepsilon_{xx}^2 + 2\nu_{zx}E_x \varepsilon_{xx}\varepsilon_{zz} + \frac{1}{2}(1-\nu_{xy})E_z\varepsilon_{zz}^2 \right].
\end{equation}
Taking small $\hat{\alpha} \equiv \alpha-1 \simeq \varepsilon_{xx}-\varepsilon_{zz}$ and $\hat{\beta} \equiv \beta -1 \simeq 2 \varepsilon_{xx}+\varepsilon_{zz}$, we can take $\varepsilon_{xx} \simeq (\hat{\alpha}+\hat{\beta})/3$ and $\varepsilon_{zz} \simeq (\hat{\beta}-2 \hat{\alpha})/3$. This allows us to easily minimize $u$ with respect to $\hat{\alpha}$ (keeping $\hat{\beta}$ fixed) to obtain 
\begin{equation}\label{eq:alpha-linear-beta}
    \alpha-1 \simeq m(\beta - 1)
\end{equation}
where
\begin{equation}\label{eq:alpha-linear-beta-slope}
    m = \frac{(1-\nu_{xy})E_z/E_x - (1-\nu_{zx})}{2(1-\nu_{xy})E_z/E_x + (1-4\nu_{zx})}.
\end{equation}
If the elasticity is isotropic (with $E_z/E_x=1$ and $\nu_{xy}=\nu_{zx}$) then the swelling is too ($m=0$) --- so that the observation of swelling anisotropy ($m>0$) implies elastic anisotropy.

We can re-express $m$ in terms of two parameters 
\begin{equation}\label{eq:slope_near_half}
    m = \frac{1}{2} + \frac{ 3(\nu_{zx}-1/2)}
        {2 (\gamma-2 \nu_{zx}^2)+(2 \nu_{zx}-1)^2}, 
\end{equation}
where $\gamma \equiv (1-\nu_{xy}) E_z/E_x$. For elastic stability we need $u>0$ in Eq.~\ref{eq:strain-potential-energy-density}, which implies $1-\nu_{xy}-2 \nu_{zx} \nu_{xz}>0$ (Eqn.~11 of \cite{Lempriere:1968}). From the symmetric compliance we have $\nu_{zx}/E_z=\nu_{xz}/E_x$, which then gives $\gamma > 2 \nu_{zx}^2$. As a result, the denominator in Eq.~\ref{eq:slope_near_half} must be strictly positive. From the previous section we expect, and observe, that the slope $m<1/2$. This implies that the axial Poisson ratio $\nu_{zx}<1/2$.  

Typically, $\nu_{zx}$ is directly measured in  uniaxial extension experiments (see Discussion). Solving for the unknown $\gamma$ in Eq.~\ref{eq:slope_near_half} we then have 
\begin{equation} \label{gamma}
\gamma = \frac{1}{2} + \frac{1+4m}{1-2m} \left(\frac{1}{2}-\nu_{zx}\right). 
\end{equation}
Since we expect $\nu_{zx}<1/2$ and $m<1/2$, we then also expect $\gamma>1/2$.  This satisfies the stability requirement $\gamma>2 \nu_{zx}^2$ when $\nu_{zx}<1/2$. 

\section{Discussion}
\begin{figure}[t]
    \centering
    \includegraphics[scale=0.7]{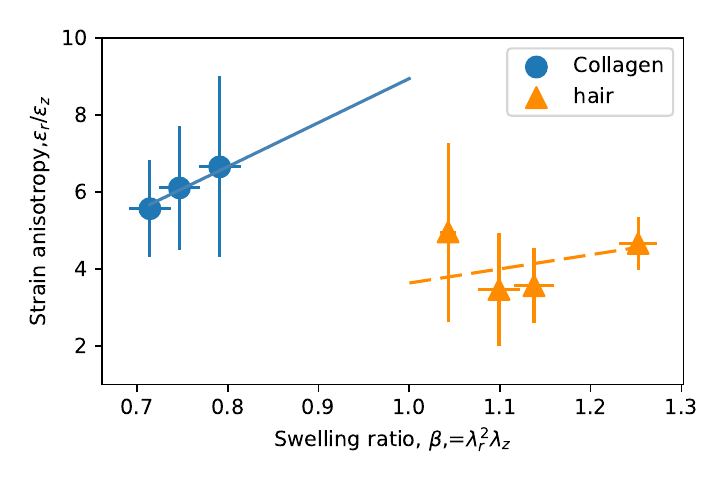}
    \caption{Strain anisotropy vs. swelling ratio. Using the same data as Fig.~\ref{fig:alpha vs beta}, we plot the strain anisotropy $\varepsilon_r / \varepsilon_z$ vs the swelling ratio $\beta = \lambda_r^2 \lambda_z = V / V_0$ --- for both hair and collagen as indicated. Note that $\epsilon = \lambda -1$. The lines assume Eqs.~\ref{eq:alpha vs theta}-\ref{eq:beta vs theta} together with the exact fit values of $A$ and $B$.}
    \label{fig:strain ratio vs swell}
\end{figure}

We have parameterized the relationship between water content and anisotropic swelling in two protein fibers --- hair shafts and collagen fibrils. We have shown that available experimental data exhibits an approximately linear dependence between shape measures and the water content, with remarkably similar behavior between hair shafts and collagen fibrils.  A plot of the strain anisotropy $\epsilon_r/\epsilon_z$ (where $\epsilon= \lambda-1$) highlights the large ${\sim}5$-fold strain anisotropies that are observed (Fig.~\ref{fig:strain ratio vs swell}). It also highlights the considerable experimental uncertainty that remains. 

We have connected the anisotropic swelling to anisotropic elasticity, under the assumption that swelling due to hydration chooses an anisotropic shape that minimizes the elastic energy. Our results constrain the elastic constants. From assuming that increasing water increases both the radial and axial deformations, we expect that $m<1/2$ --- which we observe. This in turn constrains the well-measured axial Poisson's ratio $\nu_{zx}<1/2$. These additionally constrain the product $\gamma \equiv (1-\nu_{xy})E_z/E_x >1/2$ --- which also satisfies elastic stability. While $\nu_{xy}$ has not been measured, $\nu_{zx}$ has been. 

For hair shafts, $\nu_{zx} = 0.37 \pm 0.05$ \cite{Lee:2013}, and using $m_\text{hair} = 0.32 \pm 0.03$ we  estimate that $\gamma_\text{hair} = 1.3 \pm 0.4$. This satisfies $\gamma>1/2$. With the further assumption that neither hair shafts nor collagen fibrils are radially auxetic (i.e. that $\nu_{xy} \geq 0$) we expect that $E_z/E_x \geq \gamma$ i.e. $E_z/E_x \geq 1.3 \pm 0.4$ in hair. The mechanical anisotropy of hair shafts has been recently measured \cite{Breakspear:2018}, albeit the analysis assumed $\nu_{xy}=1/2$. For hair cortex, they found $E_z/E_x \simeq 1.95$ at small relative humidity  \cite{Breakspear:2018} --  in agreement with our bound. Though the uncertainties are large, we can use this to crudely estimate $\nu_{xy} = 1-\gamma/(E_z/E_x) \approx 0.33$ for dry hair. Increased precision of the experiments could improve and constrain this estimate. 

For collagen, recent x-ray studies of bovine pericardial tissue estimated $\nu_{zx} = 2.1 \pm 0.7$ \cite{Wells:2015}. A similar value ($\nu_{xz} \approx 1.9$) can be extracted from AFM studies of tendon \cite{Rigozzi:2013, Wells:2015}. These substantially violate our bound $\nu_{zx}<1/2$. However, both of these studies involved estimates of fibril properties from tissue --- which likely violates the transversely isotropic symmetry assumed in our treatment. Studies of individual (isolated from rat tail tendon) collagen fibrils instead report $\nu_{zx} \approx 0$  (see e.g. Fig.~S11a of \cite{Peacock:2019}) --- which satisfies our bound. Using $\nu_{zx}=0$, we estimate $\gamma \simeq 10$, and we bound $E_z/E_x \geq 10$.  Brillouin scattering experiments have estimated the elastic anisotropy of wet collagen tissue and extracted $E_z/E_x \approx 1.8$ \cite{Cusack:1979} --- recent measurements of dry collagen tissue obtain all elastic constants and a similar small ratio \cite{Edginton:2016}. In contrast,  mechanical study of individual collagen fibrils has determined both $E_x \approx \SI{15}{MPa}$ by AFM nanoindentation and $E_z \approx \SI{0.4}{GPa} - \SI{0.8}{GPa}$ by AFM tensile tests \cite{Andriotis:2018, Andriotis:2023} --- resulting in $E_z/E_x \approx 27 - 54$, consistent with our bounds. We can use these to estimate $\nu_{xy} = 1-\gamma/(E_z/E_x) \approx 0.6-0.8$ from single-fibril studies.

Both hair shafts and collagen fibrils have anisotropic structures with similar anisotropic response to hydration (see Fig.~\ref{fig:alpha vs beta}). Nevertheless, the microscopic mechanisms for this response is thought to differ. For collagen fibrils, axial deformations are thought to be minimized by the sliding of collagen molecules past each other~\cite{Haverkamp:2022}. In contrast, intermediate filaments in hair shafts do not themselves restructure by sliding, but rather they guide the anisotropic swelling of the surrounding amorphous matrix~\cite{Robbins:2012, Murthy:2019}. These different microstructural models should, in principle, lead to different anisotropic properties of these materials --- as is seen in our estimates of the elastic parameters.

Elastic parameters vary with humidity and hence water content (see e.g. \cite{Breakspear:2018}) --- consistent with the enormous differences between hydrated and dry samples (see e.g. \cite{Andriotis:2023}). For hair, the elastic anisotropy $E_z/E_x$ decreases with increasing water content to $\simeq 0.9$ \cite{Breakspear:2018}. Our elasticity treatment applies for small (linear) deformations with respect to any reference water content, and so should still apply if we take $d \alpha \simeq m \; d\beta$ in Eq.~\ref{eq:alpha-linear-beta}, for small changes $d \alpha$ and $d \beta$. In other words $m = d \alpha/d \beta$, the local slope in Fig.~\ref{fig:alpha vs beta}. From $m<1/2$, we also expect the slope $d \alpha/d\beta<1/2$. From Eq.~\ref{gamma} the slope determines the combination $\gamma = (1-\nu_{xy}) E_z/E_x$, but presumably $\nu_{xy}$ is also hydration dependent in hair. The approximate linearity of $\alpha$ vs $\beta$ in Fig.~\ref{fig:alpha vs beta} indicates that the net hydration dependence of $\gamma$ may not be strong in either hair or collagen.

We expect that our results would also apply to other anisotropic biomaterials, bio-mimetic materials, and synthetic  materials. For example, Mredha et al.~\cite{Mredha:2017} created a millimeter diameter bio-mimetic ``hair'' by crosslinking a water soluble polymer solution containing aligned collagen fibrils.  They achieved hydrogels with axial deformations after hydration that ranged from 1.04 to 1.29, and radial deformations from 1.18 to 1.7. Despite using very different chemistries from hair, all formulations showed significant swelling strain anisotropy between 2 and 5. Our results indicate that considerable elastic anisotropy would be expected as well.

We have assumed a straight and radially symmetric geometry, which is not generically true for hair shafts \cite{Wolfram:2003}. In particular curly hair has asymmetric structure in cross-section \cite{Thibaut:2007}, as with wool \cite{Munro:1999}. Hair is also approximately elliptical in cross section \cite{Wolfram:2003}, though the data we have used is from hair with small cross-sectional eccentricity \cite{Stam:1952}. The structural complexity (inhomogeneity) of single hair shafts \cite{Wolfram:2003} may also lead to more complex behavior. These variations would complicate our mechanical analysis, and presumably also swelling behavior, though probably in interesting ways.  Whether water content controlled by solution conditions (see e.g. \cite{Grant:2009, Masic:2015, Haverkamp:2022}) or cross-linking density (see e.g. \cite{Miles:2005, Andriotis:2019, Vaez:2023}) has the same phenomenology as water content controlled by humidity also remains an open question experimentally, since our swelling results are based on studies that have only varied relative humidity.  We have shown how linear anisotropic hydration would be related to linear anisotropic elasticity given our symmetry assumptions. Previous studies of hair shafts and collagen fibrils should be systematically revisited to validate how shape and mechanical properties depend on humidity and water content. Similar studies for other isolated biomaterial fibers would also test and extend our results. 

\section*{Conflicts of interest}
There are no conflicts to declare.
\section*{Acknowledgements}
We thank the Natural Sciences and Engineering Research Council of Canada (NSERC) for operating Grants RGPIN-2024-04192 (LK) and RGPIN-2019-05888 (ADR).

\bibliographystyle{elsarticle-num} 
\bibliography{ref}
\end{document}